\documentclass[12pt]{article}
\usepackage{axodraw,bbold}

\catcode`@=12
\begin{document}
\thispagestyle{empty}
\begin{flushright} 
UCRHEP-T397\\ 
August 2005\
\end{flushright}
\vspace{0.5in}
\begin{center}
{\LARGE	\bf Neutrino Mass Matrix from $S_4$ Symmetry\\}
\vspace{1.5in}
{\bf Ernest Ma\\}
\vspace{0.2in}
{\sl Physics Department, University of California, Riverside, 
California 92521, USA,\\
and Institute for Particle Physics Phenomenology, Department of Physics,\\ 
University of Durham, Durham, DH1 3LE, UK\\}
\vspace{1.5in}
\end{center}

\begin{abstract}\
The cubic symmetry $S_4$ contains $A_4$ and $S_3$, both of which have 
been used to study neutrino mass matrices.  Using $S_4$ as the family 
symmetry of a complete supersymmetric theory of leptons, it is shown how 
the requirement of breaking $S_4$ at the seesaw scale without breaking 
supersymmetry enforces a special form of the neutrino mass matrix which 
exhibits maximal $\nu_\mu-\nu_\tau$ mixing as well as zero $U_{e3}$. 
In addition, $(\nu_e + \nu_\mu + \nu_\tau)/\sqrt 3$ is naturally close to 
being  a mass eigenstate, thus predicting $\tan^2 \theta_{12}$ to be near 
but not equal to 1/2.
\end{abstract}

\newpage
\baselineskip 24pt

In atmospheric neutrino oscllation data \cite{atm}, the persistence of 
maximal $\nu_\mu-\nu_\tau$ mixing has raised the important theoretical 
question of whether it is due to an underlying symmetry.  The naive 
response is that it is due to the exchange of $\nu_\mu$ with 
$\nu_\tau$, but since $\nu_\mu$ and $\nu_\tau$ are respective members of 
the $SU(2)_L$ doublets $(\nu_\mu,\mu)$ and $(\nu_\tau,\tau)$, this symmetry 
automatically implies the exchange of $\mu$ with $\tau$. As such, it cannot be 
sustained in the full Lagrangian, because the assumed diagonality of the 
charged-lepton mass matrix (witn $m_\mu \neq m_\tau$) is then lost.

A related issue is whether or not $U_{e3}=0$.  If $\nu_\mu-\nu_\tau$ 
exchange were a legitimate symmetry, then this would also be ``predicted''. 
On the other hand, $U_{e3}=0$ by itself can be supported by a different 
symmetry of the full Lagrangian \cite{cfm04}, although the latter is generally 
unable to shed any light on how maximal $\nu_\mu-\nu_\tau$ mixing could be 
achieved.

With the implementation of non-Abelian discrete symmetries such as $S_3$ 
\cite{s3}, $D_4$ \cite{d4}, and $A_4$ \cite{a4}, as well as $Q_4$ \cite{q4} 
and $Q_6$ \cite{q6}, it is indeed possible to have both maximal 
$\nu_\mu-\nu_\tau$ mixing and zero $U_{e3}$, as well as a prediction of the 
mixing angle in solar neutrino oscillations \cite{sno} in some cases.  
However, specific {\it ad hoc} assumptions regarding the symmetry breaking 
sector must be made, usually with the addition of arbitrary auxiliary Abelian 
discrete symmetries \cite{af05}.

In this paper, the problem is solved by invoking a very simple requirement.  
The complete theory of leptons is assumed to be supersymmetric with $S_4$ as 
its family symmetry.  Neutrino masses are assumed to come from the canonical 
seesaw mechanism \cite{seesaw} with heavy singlet neutral fermions $N$. The 
key is to allow $S_4$ to be broken by the Majorana mass matrix of $N$ at the 
seesaw scale, but not the supersymmetry. This requirement then fixes the 
pattern of symmetry breaking and thus the form of ${\cal M}_N$, and 
subsequently also ${\cal M}_\nu$, as shown below.

The group of the permutation of four objects is $S_4$ \cite{ysp}. It is also 
the symmetry group of the hexahedron, i.e. the cube, one of five (and only 
five) perfect geometric solids, which was identified by Plato with the 
element ``earth'' \cite{plato}. It has 24 elements divided into 5 equivalence 
classes, with \underline{1}, \underline{1}$'$, \underline{2}, \underline{3}, 
and \underline{3}$'$ as its 5 irreducible representations.  Its character 
table is given below. 

\begin{table}[htb]
\caption{Character table of $S_4$.}
\begin{center}
\begin{tabular}{|c|c|c|c|c|c|c|c|}
\hline 
Class & $n$ & $h$ & $\chi_1$ & $\chi_{1'}$ & $\chi_2$ & $\chi_3$ & $\chi_{3'}$ 
\\ 
\hline
$C_1$ & 1 & 1 & 1 & 1 & 2 & 3 & 3 \\ 
$C_2$ & 3 & 2 & 1 & 1 & 2 & --1 & --1 \\ 
$C_3$ & 8 & 3 & 1 & 1 & --1 & 0 & 0 \\ 
$C_4$ & 6 & 4 & 1 & --1 & 0 & --1 & 1 \\ 
$C_5$ & 6 & 2 & 1 & --1 & 0 & 1 & --1 \\ 
\hline
\end{tabular}
\end{center}
\end{table}

The two three-dimensional representations differ only in the signs of their 
$C_4$ and $C_5$ matrices.  Their group multiplication rules are similar to 
those of $A_4$ \cite{mr01}, namely
\begin{eqnarray}
\underline{3} \times \underline{3} &=& \underline{1} + \underline{2} + 
\underline{3}_S + \underline{3}'_A, \\
\underline{3}' \times \underline{3}' &=& \underline{1} + \underline{2} + 
\underline{3}_S + \underline{3}'_A, \\
\underline{3} \times \underline{3}' &=& \underline{1}' + \underline{2} + 
\underline{3}'_S + \underline{3}_A,
\end{eqnarray}
where the subscripts $S$ and $A$ refer to their symmetric and antisymmetric 
product combinations respectively.  The two-dimensional representation 
behaves exactly as its $S_3$ counterpart, namely
\begin{equation}
\underline{2} \times \underline{2} = \underline{1} + \underline{1}' + 
\underline{2}.
\end{equation}

The three $N$'s are assigned to the \underline{3} representation. To obtain 
a nontrivial ${\cal M}_N$, the singlet Higgs superfields $\sigma_{1,2,3} \sim 
\underline{3}$ and $\zeta_{1,2} \sim \underline{2}$ are assumed. 
Consequently in the $N_{1,2,3}$ basis,
\begin{equation}
{\cal M}_N = \pmatrix{A + f(\langle \zeta_2 \rangle + \langle \zeta_1 \rangle) 
& h \langle \sigma_3 \rangle & h \langle \sigma_2 \rangle \cr 
h \langle \sigma_3 \rangle & A + f(\omega \langle \zeta_2 \rangle + 
\omega^2 \langle \zeta_1 \rangle) & h \langle \sigma_1 \rangle \cr 
h \langle \sigma_2 \rangle & h \langle \sigma_1 \rangle & A + f(\omega^2 
\langle \zeta_2 \rangle + \omega \langle \zeta_1 \rangle)},
\end{equation}
where $\omega=\exp(2 \pi i/3)$.  The most general $S_4$-invariant 
superpotential of $\sigma$ and $\zeta$ is given by
\begin{eqnarray}
W &=& {1 \over 2} M (\sigma_1 \sigma_1 + \sigma_2 \sigma_2 + 
\sigma_3 \sigma_3) + \lambda \sigma_1 \sigma_2 \sigma_3 + m \zeta_1 \zeta_2 + 
{1 \over 3} \rho (\zeta_1 \zeta_1 \zeta_1 + \zeta_2 \zeta_2 \zeta_2) \nonumber 
\\ && + {1 \over 2} \kappa ( \sigma_1 \sigma_1 + \omega \sigma_2 \sigma_2 + 
\omega^2 \sigma_3 \sigma_3) \zeta_2 + {1 \over 2} \kappa ( \sigma_1 \sigma_1 
+ \omega^2 \sigma_2 \sigma_2 + \omega \sigma_3 \sigma_3) \zeta_1.
\end{eqnarray}
The resulting scalar potential is then
\begin{eqnarray}
V &=& |M \sigma_1 + \lambda \sigma_2 \sigma_3 + \kappa \sigma_1 (\zeta_2 + 
\zeta_1)|^2 \nonumber \\ 
&+& |M \sigma_2 + \lambda \sigma_3 \sigma_1 + \kappa \sigma_2 (\omega \zeta_2 
+ \omega^2 \zeta_1)|^2 \nonumber \\ 
&+& |M \sigma_3 + \lambda \sigma_1 \sigma_2 + \kappa \sigma_3 (\omega^2 
\zeta_2 + \omega \zeta_1)|^2 \nonumber \\ 
&+& |m \zeta_1 + \rho \zeta_2 \zeta_2 + {1 \over 2} \kappa ( \sigma_1 \sigma_1 
+ \omega \sigma_2 \sigma_2 + \omega^2 \sigma_3 \sigma_3)|^2 \nonumber \\ 
&+& |m \zeta_2 + \rho \zeta_1 \zeta_1 + {1 \over 2} \kappa ( \sigma_1 \sigma_1 
+ \omega^2 \sigma_2 \sigma_2 + \omega \sigma_3 \sigma_3)|^2.
\end{eqnarray}
For supersymmetry to be unbroken, $V_{min}=0$ is required.  This is possible 
only if  $\langle \zeta_1 \rangle = \langle \zeta_2 \rangle$ and  $\langle 
\sigma_2 \rangle = \langle \sigma_3 \rangle$, for which
\begin{eqnarray}
&& M \langle \sigma_1 \rangle + \lambda \langle \sigma_2 \rangle^2 + 2 \kappa 
\langle \sigma_1 \rangle \langle \zeta_1 \rangle = 0, \\ 
&& M + \lambda \langle \sigma_1 \rangle - \kappa \langle \zeta_1 \rangle = 0, 
\\ 
&& m \langle \zeta_1 \rangle + \rho \langle \zeta_1 \rangle^2 + {1 \over 2} 
\kappa (\langle \sigma_1 \rangle^2 - \langle \sigma_2 \rangle^2) = 0,
\end{eqnarray}
where $\omega + \omega^2 = -1$ has been used. Thus ${\cal M}_N$ is fixed to 
be of the form
\begin{equation}
{\cal M}_N = \pmatrix{A+2B & C & C \cr C & A-B & D \cr C & D & A-B},
\end{equation}
where $B=f\langle \zeta_1 \rangle$, $C=h\langle \sigma_2 \rangle$, and 
$D=h\langle \sigma_1 \rangle$.

The residual symmetry of the theory is then $Z_2$, under which $(N_2-N_3)$, 
$(\sigma_2-\sigma_3)$, $(\zeta_1-\zeta_2)$ are odd, and $N_1$, $(N_2+N_3)$, 
$\sigma_1$, $(\sigma_2+\sigma_3)$, $(\zeta_1+\zeta_2)$ are even.  If $\langle 
\zeta_{1,2} \rangle = 0$, then $B=0$, $C=D$, and the residual symmetry is 
$S_3$, under which $(N_1 + N_2 + N_3)$, $(\sigma_1 + \sigma_2 + \sigma_3) 
\sim \underline{1}$; $(N_1 + \omega N_2 + \omega^2 N_3, N_1 + \omega^2 N_2 
+ \omega N_3), (\sigma_1 + \omega \sigma_2 + \omega^2 \sigma_3, \sigma_1 + 
\omega^2 \sigma_2 + \omega \sigma_3), (\zeta_1, \zeta_2) \sim \underline{2}$.  
The $Z_2$ (and the approximate $S_3$) symmetry of ${\cal M}_N$ 
is preserved in ${\cal M}_N^{-1}$.

Consider now the leptons $(\nu_i,l_i)$ and $l^c_j$ of the Standard Model. 
They are also assigned to the \underline{3} representation of $S_4$. 
As for the Higgs doublet superfields, they are assumed to be \underline{1} + 
\underline{2}, with one set coupling $(\nu_i,l_i)$ to $l^c_j$ and 
the other coupling $(\nu_i,l_i)$ to $N_j$, as in any supersymmetric 
extension of the Standard Model.  Nonzero vacuum expectation values of 
the scalar components of these Higgs superfields break the electroweak 
symmetry, but because of their $S_4$ structure, the $l_i l^c_j$ mass 
matrix is diagonal, as well as that of $\nu_i N_j$. In particular, 
\begin{equation}
\pmatrix{m_e \cr m_\mu \cr m_\tau} = \pmatrix{1 & 1 & 1 \cr 1 & 
\omega & \omega^2 \cr 1 & \omega^2 & \omega} 
\pmatrix{y_1 v_1 \cr y_2 v_2 \cr y_2 v_3},
\end{equation}
where $v_1$ is the vacuum expectation value of the \underline{1} 
representation, $v_{2,3}$ are those of the \underline{2}, and $y_{1,2}$ 
are their respective Yukawa couplings.  In the $\nu N$ sector, the vacuum 
expectation values of the corresponding Higgs \underline{2} representation 
are assumed to be zero, thus $m_1=m_2=m_3(=m_D)$ for all the Dirac neutrino 
masses.  Using the seesaw mechanism, the observed Majorana neutrino mass 
matrix in the $\nu_{e,\mu,\tau}$ basis is then given by
\begin{equation}
{\cal M}_\nu = -{\cal M}_D {\cal M}_N^{-1} {\cal M}_D^T = \pmatrix{a+2b & 
c & c \cr c & a-b & d \cr c & d & a-b},
\end{equation}
where
\begin{eqnarray}
a &=& [-A^2 + B^2 + (2 C^2 + D^2)/3] m_D^2/det{\cal M}_N, \\ 
b &=& [B(A-B) - (C^2-D^2)/3] m_D^2/det{\cal M}_N, \\
c &=& C(A-B-D) m_D^2/det{\cal M}_N, \\
d &=& [D(A+2B) - C^2] m_D^2/det{\cal M}_N, \\
det{\cal M}_N &=& (A-B-D)[(A+2B)(A-B+D)-2C^2].
\end{eqnarray}
This form of ${\cal M}_\nu$ is precisely that advocated in Ref.~\cite{m02} 
on purely phenomenological grounds.  As expected, the $S_3$ limit is obtained 
if $B=0$ and $C=D$, for which $b=0$ and $c=d$, resulting in $(\nu_e + \nu_\mu 
+ \nu_\tau)/\sqrt 3$ as a mass eigenstate.  Note that $S_4$ as well as the 
residual $Z_2$ symmetry are assumed broken softly at the scale of 
supersymmetry breaking.  This allows the electroweak vacuum expectation 
values to be chosen as they are.

In the basis $[\nu_e, (\nu_\mu + \nu_\tau)/\sqrt 2, (-\nu_\mu + \nu_\tau)
/\sqrt 2]$, the neutrino mass matrix of Eq.~(13) becomes
\begin{equation}
{\cal M}_\nu = \pmatrix{a+2b & \sqrt 2 c & 0 \cr \sqrt 2 c & a-b +d & 0 
\cr 0 & 0 & a-b-d},
\end{equation}
which exhibits maximal $\nu_\mu-\nu_\tau$ mixing and zero $U_{e3}$ as 
advertised.  The mixing angle of the $2 \times 2$ submatrix (for all 
parameters real) can be simply read off as
\begin{equation}
\tan 2 \theta_{12} = {2 \sqrt 2 c \over d - 3b} = {2 \sqrt 2 c \over 
c - 3b - (c-d)},
\end{equation}
which reduces to $2 \sqrt 2$ in the $S_3$ limit of $b=0$ and $c=d$. 
This would imply $\tan^2 \theta_{12} = 1/2$, resulting in a mixing 
pattern proposed some time ago \cite{hps}.  However in this limit, 
$\Delta m^2_{atm}$ vanishes as well \cite{cfm05}. Therefore, 
$\tan^2 \theta_{12}$ should not be equal to 1/2, but since it is 
natural for $b$ and $c-d$ to be small compared to $c$, its deviation 
from 1/2 is expected to be small.  For example,
\begin{equation}
(3b+c-d)/c = -0.15 ~\Longrightarrow~ \tan^2 \theta_{12} = 0.45,
\end{equation}
in excellent agreement with data \cite{sno}. 

The limit $\Delta m^2_{sol}=0$ implies $2a+b+d=0$, and
\begin{eqnarray}
\Delta m^2_{atm} &\equiv& m_3^2 - (m_2^2+m_1^2)/2 = (a-b-d)^2 - (d-3b)^2/4 - 
2 c^2 \nonumber \\ &=& 6bd - 2(c^2 - d^2) \simeq [6b - 4(c - d)]c,
\end{eqnarray}
which can be either positive or negative, corresponding to a normal or 
inverted ordering of neutrino masses respectively.  If $c-d=0$, then $\tan^2 
\theta_{12} < 1/2$ implies an inverted ordering as in the models of 
Ref.~\cite{cfm05}.  It is also possible to set $B=0$ alone by choosing $f=0$ 
in Eq.~(5), so that Eq.~(20) becomes
\begin{equation}
\tan 2 \theta_{12} = {2 \sqrt 2 C \over D},
\end{equation}
and $\Delta m^2_{sol} = 0$ implies
\begin{equation}
\Delta m^2_{atm} \simeq 9 (C-D)C^3(m_D^2/det{\cal M}_N)^2.
\end{equation}
Thus $\tan^2 \theta_{12}<1/2$ would also imply an inverted ordering of 
neutrino masses in this case.

The effective neutrino mass $m_{ee}$ measured in neutrinoless double beta 
decay is simply given by the magnitude of the $\nu_e \nu_e$ entry of 
${\cal M}_\nu$, i.e. $|a+2b|$.  Using Eqs.~(21), (22), and $|\Delta m^2_{atm}| 
= 2.5 \times 10^{-3}$ eV$^2$, it is in the range 0.10 eV for $c-d=0$ (inverted 
ordering) and 0.05 eV (normal ordering) for $b=0$.

Returning to the condition $2a+b+d=0$ for $\Delta m^2_{sol}=0$, under 
which ${\cal M}_\nu$ of Eq.~(13) becomes
\begin{equation}
{\cal M}_\nu = \pmatrix{a+2b & c & c \cr c & a-b & -2a-b \cr c & -2a-b & a-b},
\end{equation}
it should be noted that this form is invariant under the transformation 
\cite{m03}
\begin{equation}
U {\cal M}_\nu U^T = {\cal M}_\nu,
\end{equation}
where
\begin{equation}
U = \pmatrix{i 2 \sqrt 2 /3 & i/3 \sqrt 2 & i/3 \sqrt 2 \cr i/3 \sqrt 2 & 
1/2 - i \sqrt 2/3 & -1/2 - i \sqrt 2/3 \cr i/3 \sqrt 2 & -1/2 - i \sqrt 2/3 
& 1/2 - i \sqrt 2/3}, ~~~ U^2 = \pmatrix{-1 & 0 & 0 \cr 0 & 0 & -1 \cr 
0 & -1 & 0},
\end{equation}
if $c=-2a-4b$, for which $\tan^2 \theta_{12} = 1/2$.  This limit of 
${\cal M}_\nu$ corresponds thus to a $Z_4$ symmetry.

In conclusion, the family symmetry $S_4$ has been advocated as the origin 
of the observed pattern of neutrino mixing in the context of a complete 
supersymmetric theory.  The key is the requirement that supersymmetry is 
unbroken at the seesaw scale where $S_4$ is broken.  This fixes the 
pattern of $S_4$ breaking to retain a residual $Z_2$ (as well as an 
approximate $S_3$) symmetry in the Majorana mass matrix of the heavy 
singlet neutrinos.  At the much lower scale of supersymmetry 
breaking, $S_4$ as well as $Z_2$ are allowed to be broken by soft terms. 
However, the $S_4$-invariant Yukawa terms ensure that the charged-lepton 
mass matrix is diagonal, and the neutrino Dirac masses are equal. This 
results in a Majorana neutrino mass matrix which exhibits maximal 
$\nu_\mu-\nu_\tau$ mixing and zero $U_{e3}$.  In the $S_3$ limit, 
$(\nu_e+\nu_\mu+\nu_\tau)/\sqrt 3$ is a mass eigenstate, thus predicting 
$\tan^2 \theta_{12}=1/2$, but $\Delta m^2_{atm}=0$ as well.  Allowing 
a small deviation from this limit can result in $\tan^2 \theta_{12}=0.45$ 
and a nonzero $\Delta m^2_{atm}$, in excellent agreement with data.

I thank Michele Frigerio for an important comment. This work was supported 
in part by the U.~S.~Department of Energy under Grant No. DE-FG03-94ER40837.

\newpage
\noindent {\bf Appendix} The matrices of the \underline{3} representation 
of $S_4$ are given by
\begin{eqnarray}
C_1 &:& \pmatrix{1 & 0 & 0 \cr 0 & 1 & 0 \cr 0 & 0 & 1}, \\ 
C_2 &:& \pmatrix{1 & 0 & 0 \cr 0 & -1 & 0 \cr 0 & 0 & -1}, 
\pmatrix{-1 & 0 & 0 \cr 0 & 1 & 0 \cr 0 & 0 & -1}, 
\pmatrix{-1 & 0 & 0 \cr 0 & -1 & 0 \cr 0 & 0 & 1}, \\ 
C_3 &:& \pmatrix{0 & 1 & 0 \cr 0 & 0 & 1 \cr 1 & 0 & 0}, 
\pmatrix{0 & 1 & 0 \cr 0 & 0 & -1 \cr -1 & 0 & 0}, 
\pmatrix{0 & -1 & 0 \cr 0 & 0 & 1 \cr -1 & 0 & 0}, 
\pmatrix{0 & -1 & 0 \cr 0 & 0 & -1 \cr 1 & 0 & 0}, \nonumber \\
&& \pmatrix{0 & 0 & 1 \cr 1 & 0 & 0 \cr 0 & 1 & 0}, 
\pmatrix{0 & 0 & 1 \cr -1 & 0 & 0 \cr 0 & -1 & 0}, 
\pmatrix{0 & 0 & -1 \cr 1 & 0 & 0 \cr 0 & -1 & 0}, 
\pmatrix{0 & 0 & -1 \cr -1 & 0 & 0 \cr 0 & 1 & 0}, \\ 
C_4 &:& \pmatrix{-1 & 0 & 0 \cr 0 & 0 & 1 \cr 0 & -1 & 0}, 
\pmatrix{0 & 0 & -1 \cr 0 & -1 & 0 \cr 1 & 0 & 0}, 
\pmatrix{0 & 1 & 0 \cr -1 & 0 & 0 \cr 0 & 0 & -1}, \nonumber \\ 
&& \pmatrix{-1 & 0 & 0 \cr 0 & 0 & -1 \cr 0 & 1 & 0}, 
\pmatrix{0 & 0 & 1 \cr 0 & -1 & 0 \cr -1 & 0 & 0}, 
\pmatrix{0 & -1 & 0 \cr 1 & 0 & 0 \cr 0 & 0 & -1}, \\ 
C_5 &:& \pmatrix{1 & 0 & 0 \cr 0 & 0 & 1 \cr 0 & 1 & 0}, 
\pmatrix{0 & 0 & 1 \cr 0 & 1 & 0 \cr 1 & 0 & 0}, 
\pmatrix{0 & 1 & 0 \cr 1 & 0 & 0 \cr 0 & 0 & 1}, \nonumber \\ 
&& \pmatrix{1 & 0 & 0 \cr 0 & 0 & -1 \cr 0 & -1 & 0}, 
\pmatrix{0 & 0 & -1 \cr 0 & 1 & 0 \cr -1 & 0 & 0}, 
\pmatrix{0 & -1 & 0 \cr -1 & 0 & 0 \cr 0 & 0 & 1}. 
\end{eqnarray}
The matrices of the \underline{3}$'$ representation are the same 
as those of \underline{3} for $C_{1,2,3}$ and oppposite in sign for $C_{4,5}$.
Those of the \underline{2} representation are the same as in $S_3$, i.e.
\begin{eqnarray}
C_{1,2} &:& \pmatrix{1 & 0 \cr 0 & 1}, \\ 
C_3 &:& \pmatrix{\omega & 0 \cr 0 & \omega^2}, 
\pmatrix{\omega^2 & 0 \cr 0 & \omega}, \\ 
C_{4,5} &:& \pmatrix{0 & 1 \cr 1 & 0}, 
\pmatrix{0 & \omega \cr \omega^2 & 0}, 
\pmatrix{0 & \omega^2 \cr \omega & 0}, 
\end{eqnarray}
each appearing four times.

\bibliographystyle{unsrt}

\begin{thebibliography}{99}
\bibitem{atm} Y. Ashie {\it et al.}, Phys. Rev. {\bf D71}, 112005 (2005) and 
references therein.
\bibitem{cfm04} See for example S.-L. Chen, M. Frigerio, and E. Ma, Phys. 
Lett. {\bf B612}, 29 (2005) and references therein.
\bibitem{s3} See for example J. Kubo, A. Mondragon, M. Mondargon, and E. 
Rodriguez-Jauregui, Prog. Theor. Phys. {\bf 109}, 795 (2003); S.-L. Chen, 
M. Frigerio, and E. Ma, Phys. Rev. {\bf D70}, 073008 (2004) and references 
therein.
\bibitem{d4} See for example W. Grimus, A. S. Joshipura, S. Kaneko, 
L. Lavoura, and M. Tanimoto, JHEP {\bf 0407}, 078 (2004).
\bibitem{a4} See for example E. Ma, hep-ph/0508099 and references therein.
\bibitem{q4} M. Frigerio, S. Kaneko, E. Ma, and M. Tanimoto, Phys. Rev. 
{\bf D71}, 011901(R) (2005). The quaternion group is denoted here as $Q_8$, 
but is called $Q_4$ in the convention of Ref.~\cite{q6}.
\bibitem{q6} K. S. Babu and J. Kubo, Phys. Rev. {\bf D71}, 056006 (2005); 
J. Kubo, hep-ph/0506043.
\bibitem{sno} B. Aharmim {\it et al.}, nucl-ex/0502021.
\bibitem{af05} See for example G. Altarelli and F. Feruglio, Nucl.Phys. 
{\bf B720}, 64 (2005); K. S. Babu and X.-G. He, hep-ph/0507217.
\bibitem{seesaw} M. Gell-Mann, P. Ramond, and R. Slansky, in 
{\em Supergravity}, edited by P. van Nieuwenhuizen and D. Z. Freedman 
(North-Holland, Amsterdam, 1979), p.~315; T. Yanagida, in {\em Proceedings 
of the Workshop on the Unified Theory and the Baryon Number in the Universe}, 
edited by O. Sawada and A. Sugamoto (KEK Report No.~79-18, Tsukuba, Japan, 
1979), p.~95; R. N. Mohapatra and G. Senjanovic, Phys. Rev. Lett. {\bf 44}, 
912 (1980).
\bibitem{ysp} Y. Yamanaka, H. Sugawara, and S. Pakvasa, Phys. Rev. {\bf D25}, 
1895 (1982); Erratum: ibid. {\bf D29}, 2135 (1984).
\bibitem{plato} E. Ma, Mod. Phys. Lett. {\bf A17}, 2361 (2002).
\bibitem{mr01} E. Ma and G. Rajasekaran, Phys. Rev. {\bf D64}, 113012 (2001).
\bibitem{m02} E. Ma, Phys. Rev. {\bf D66}, 117301 (2002).
\bibitem{hps} P. F. Harrison, D. H. Perkins, and W. G. Scott, Phys. Lett. 
{\bf B530}, 167 (2002).
\bibitem{cfm05} S.-L. Chen, M. Frigerio, and E. Ma, hep-ph/0504181, M. Hirsch, 
A. Villanova del Moral, J. W. F. Valle, and E. Ma, hep-ph/0507148.
\bibitem{m03} E. Ma, Phys. Rev. Lett. {\bf 90}, 221802 (2003); E. Ma and G. 
Rajasekaran, Phys. Rev. {\bf D68}, 031702(R) (2003).
\end{thebibliography}

\end{document}